\documentclass[showpacs,preprintnumbers,prd,nofootinbib,floats,amssymb,floatfix]{revtex4}
\usepackage{amsmath}
\usepackage{amsfonts}
\usepackage{graphicx}
\usepackage{hyperref}
\usepackage{amsmath}
\usepackage{amstext}
 
\begin{document}
\title{Hamiltonian dynamics of an exotic action for gravity in three dimensions }
\author{Alberto Escalante}  \email{aescalan@ifuap.buap.mx}
\author{J. Manuel-Cabrera } \email{jmanuel@ifuap.buap.mx}
 \affiliation{  Instituto de F{\'i}sica, Universidad Aut\'onoma de Puebla, \\
 Apartado Postal J-48 72570, Puebla Pue., M\'exico, }
\begin{abstract}
The Hamiltonian dynamics  and the canonical covariant formalism for an exotic action in three  dimensions  are performed.  By working   with the complete  phase  space,  we report a complete Hamiltonian description   of the theory such as  the extended action, the extended Hamiltonian, the  algebra among  the constraints, the Dirac's brackets and  the correct gauge transformations.  In addition, we show that in spite of  exotic action and  tetrad gravity with a cosmological constant  give rise to the same equations of motion, they are not equivalent, in fact, we show that their corresponding Dirac's brackets are quite different. Finally, we  construct   a  gauge invariant symplectic form which in turn represents a complete Hamiltonian description of the covariant phase space.
\end{abstract}
 \date{\today}
\pacs{98.80.-k,98.80.Cq}
\preprint{}
\maketitle
\section{INTRODUCTION}
A dynamical system is characterized by means of its symmetries which constitute an important information in both the classical and the quantum context. It is well-known that the analysis of a dynamical system by means of its equations of motion implies that the phase space is not endowed with a natural or preferred symplectic structure as it has been claimed in \cite{1a, 1}, and the freedom in the choice of the symplectic structure is an important issue because it could yield different quantum formulations. Hence, in spite we have an infinite way to choose a symplectic structure for any system, the following question arises:  are there the same symmetries in two different actions sharing the same equations of motion? The answer in general is not. In fact, it has been showed that two theories sharing the same equations of motion, does not imply that the theories are equivalent even at the classical level \cite{2,3}. Nonetheless, the study of any theory  should  be carried out extending  the definition of a dynamical system by considering its equations of motion plus an action principle, thus we are in a profitable situation because the action gives us the equations
of motion and symmetries; additionally  it fixes the symplectic structure of the theory \cite{1, 4}. In this manner, in the study of the symmetries of a dynamical system must be taken into account both, the equations of motion plus an action principle \cite{3}.  Nowadays,
there exist approaches that can be used for studying  the symmetries of  any theory, as for instance, Dirac's canonical formalism and the covariant canonical method, both with their respective advantages. Dirac's canonical formalism is an elegant approach for obtaining relevant physical information of a theory under study, namely, the counting of physical degrees of freedom, the correct gauge transformations, the study of the constraints, the extended Hamiltonian and the extended action \cite{5}, all this information is the guideline to make the best progress in the analysis of quantum aspects.
On the other hand, in the covariant canonical method,  in order to describe all the relevant Hamiltonian description of the covariant phase space \cite{ 6}, we are able to identify a gauge invariant two-form, being an important step to analyze within a complete
covariant context the theory under study. Therefore, we think that the complete analysis
of any theory should be done by performing a Dirac's canonical approach  and    the
canonical covariant method, the former  because it considers the action to study   its symmetries, the latter   takes into account  the  equations of
motion in order to construct the covariant phase space. In this respect,  usually  the way to perform the Dirac  formalism is not carried out in a complete form, namely,  usually the people prefer to work on a smaller phase space context  \cite{8a, 8, 9}; this means that only those variables that occur in the action  with temporal derivative are considered as dynamical,  in general  in order to obtain a complete study one must  perform  a pure Dirac's method, this is, we need to   consider the complete set of  variables occurring in  our theory as dynamical ones.  In this respect,  we have performed a pure Dirac's canonical analysis for models as $BF$ theories, the Pontryagin invariant, topological theories,  etc.,  \cite{8, 9}  and we have reported the complete structure of the constraints defined on the full phase space, we have commented in those works,  that by performing a pure Dirac's  framework  we are able to know the  symmetries of the theory, as for instance,  gauge symmetry and the complete algebra among the constraints defined on the full phase space, fact that usually is not  possible to obtain by using a smaller phase space context.   \\
In this manner, the purpose of this paper, is to develop a complete Hamiltonian analysis of an exotic action in three dimensions. It is well-known that  Palatini's  gravity  with a cosmological constant and exotic action
yield the same equations of motion, and there are many works commenting  that this fact makes the actions classically equivalent ( see \cite{10, 11} and the references therein). However, a complete analysis of an exotic action has not been performed, and  therefore the complete symmetries of the theory are not  well known. Thus, we show in this paper  that the Dirac's brackets for the dynamical variables that define exotic action and Palatini's  gravity with a cosmological constant are different.  In fact, for the former the dynamical variables of the theory are non-commutative  and the cosmological constant can not be zero. For the Palatini action with a cosmological constant, the Dirac's brackets of dynamical variables are commutative and the cosmological constant can be taken as zero, all those ideas will be clarified along the paper. In addition we report the canonical covariant analysis of an exotic action in order to report a complete study of the theory.  By constructing a gauge invariant two form on the covariant phase space, we confirm some results obtained by means of Dirac's framework.
\section{ Hamiltonian  dynamics for Exotic action in three dimensions}
In this section, we will perform a pure Dirac's  analysis for an exotic action given by the following action \cite{11}
\begin{equation}
S[e,A]_{exotic}= \frac{1}{2} \int_M A^{IJ} \wedge d A_{IJ} + \frac{2}{3}  A^{IK} \wedge  A_{KL} \wedge A^{L}{_{I}} + \int_M \frac{\Lambda}{2} e_I\wedge D e^I ,
\label{eq21}
\end{equation}
 where $A^{IJ}=A{_{\mu}}^{IJ}dx^{\mu}$ is the Lorentz connection valued in the Lie algebra of $SO(2,1)$ and $e^{I}$ corresponds to the tetrad field or gravitational field, $\mu,\nu=0,1,2$ are spacetime indices, $x^{\mu}$ are the coordinates that label the points for the 3-dimensional spacetime manifold $M$ and $I,J=0,1,2$ are internal indices that can be raised and lowered by internal Lorentzian metric $\eta_{IJ}=(-1,1,1)$,  $D_{a} A_b{^{IJ}}= \partial_aA_b{^{IJ}}+ A_a{^{IK}}A_{b}{_K}{^{J}} +A_a{^{JK}}A_{b}{^{I}}_K $ and  $ F{^{IJ}}_{ab}=\partial_aA_b{^{IJ}} - \partial_bA_a{^{IJ}} + A_a{^{IK}}A_{b}{_K}{^{J}} - A_b{^{IK}}A_{a}{_K}{^{J}} $. \\
It is well-known  that this exotic action is the coupling of Chern-Simons theory (the first two terms on the left hand side of  (\ref{eq21}))  and the   Nieh-Yang topological term.  In the following lines we will find  an analogy among  the Nieh-Yang term and Landau's problem in the Chern-Simons quantization \cite{12}. \\
The equations of motion obtained from  (\ref{eq21}) are given by
\begin{eqnarray}
\frac{\delta S[A,e]_{exotic}}{ \delta A_{\alpha}{^{IJ}}}&:& \epsilon^{\alpha \mu\nu} R_{IJ \mu\nu}[A] - \Lambda \epsilon^{\alpha \mu\nu}e_{I \mu} e_{J \nu}=0, \nonumber \\
\frac{\delta S[A,e]_{exotic}}{ \delta e_{I \alpha}}&:&  \Lambda \epsilon^{\alpha \mu\nu} D_{\mu} e^I{_{\nu}}=0.
\label{eq4}
\end{eqnarray}
The first equations of motion refer to  Einstein's equation written in the first order  formalism, and the second  refers to  the no-torsion condition. By contracting the equations of motion with the inverse $e^d_ I$ field,  these imply that the spacetime has constant curvature equal to  $6 \Lambda$.  \\
On the other hand, we have commented above  with the terminology of a pure Dirac's
method we mean that  we will consider in the Hamiltonian framework that all the fields that define our theory are dynamical ones. It is  important to remark,  that  usually   the Hamiltonian analysis of any  theory is performed  by considering as dynamical variables only those that
occur in the Lagrangian density with temporal derivative \cite{9}. However,  the price to pay for developing the analysis on a smaller phase space  is that we cannot know the complete structure of the
constraints, their algebra and the gauge transformations defined on the  full phase space \cite{8a, 13, 10}. Hence, it is mandatory to develop a complete Hamiltonian analysis in order to report  all the relevant symmetries of the theory.\\
By performing the 2+1 decomposition of spacetime, it is assumed that the spacetime manifold is of the form $M^{3}=\Sigma\times R$, where $\Sigma$ corresponds to Cauchy's surface and $R$ represents an evolution parameter. By performing the $2+1$ decomposition, we can write the action as
\begin{equation}
S[e,A]_{exotic}= \int_M \left[ \frac{1}{2} \epsilon^{0ab} A_{0}{^{IJ}} F_{IJ ab} + \frac{1}{2} \epsilon^{0ab} A_{b}{^{IJ}} \dot{A}_{a IJ} +\frac{\Lambda}{2}\epsilon^{0ab} e_{Ib} \dot{e}{^I}_{a} + \Lambda\epsilon^{0ab} e{^I}_a D_b e_{I0} -\frac{\Lambda}{2}\epsilon^{0ab}A_{0}{^{IJ}} e_{aI} e_{bJ}\right]dx^3,
 \label{eq22}
\end{equation}
where we can identify the following  Lagrangian density
\begin{equation}
{\mathcal{L}}= \frac{1}{2} \epsilon^{0ab} A_{0}{^{IJ}} F_{IJ ab} + \frac{1}{2} \epsilon^{0ab} A_{b}{^{IJ}} \dot{A}_{a IJ} +\frac{\Lambda}{2}\epsilon^{0ab} e_{Ib} \dot{e}{^I}_{a} +\Lambda \epsilon^{0ab} e{^I}_a D_b e_{I0} -\frac{\Lambda}{2}\epsilon^{0ab} A_{0}{^{IJ}} e_{aI} e_{bJ}.
\label{eq23}
\end{equation}
Hence, by identifying our set of dynamical variables, a pure Dirac's method calls for the definition of the momenta $(\Pi^{\alpha}{_{I}},\Pi^{\alpha}{_{IJ}})$ canonically conjugate to $( e{^I}_{\alpha}, A_{\alpha} {^{IJ}})$
\begin{equation}
\Pi^{\alpha}{_{IJ}}= \frac{\delta {\mathcal{L}} }{ \delta \dot{A}_{\alpha} {^{IJ}} },  \qquad \Pi^{\alpha}{_{I}}= \frac{\delta {\mathcal{L}} }{ \delta \dot{e}{^I}_\alpha } .
\label{eq2a}
\end{equation}
The matrix elements of the Hessian
\begin{equation}
 \frac{\partial^2{\mathcal{L}} }{\partial (\partial_\mu  e^{I}_\alpha) \partial (\partial_\mu e^{I}_\beta )} , \quad \frac{\partial^2{\mathcal{L}} }{\partial( \partial_\mu  e^{I}_\alpha) \partial (\partial_\mu A_{\beta} {^{IJ}} )}, \quad \frac{\partial^2{\mathcal{L}} }{\partial (\partial_\mu A_{\alpha} {^{IJ}} ) \partial(\partial_\mu A_{\beta} {^{IJ}} ) },
\label{eq25}
\end{equation}
are identically zero,  the rank is zero, thus, we expect $18$ primary constraints. From the definition of the momenta $(\ref{eq2a})$ we identify the following $18$ primary constraints
\begin{eqnarray}
\phi{_{I}}^{0}&:=& \Pi{_{I}}^{0} \approx 0 ,\nonumber \\
\phi{_{I}}^{a}&:=& \Pi{_{I}}^{a}- \frac{\Lambda}{2}\epsilon^{0ab} e_{Ib} \approx 0 ,\nonumber \\
\phi{_{IJ}}^{0}&:=& \Pi{_{IJ}}^{0} \approx 0 ,\nonumber \\
\phi{_{IJ}}^{a}&:=& \Pi{_{IJ}}^{a}- \frac{\epsilon^{0ab}}{2}A_{bIJ}  \approx 0.
\label{eq26a}
\end{eqnarray}
The canonical Hamiltonian takes the form
\begin{equation}
H_c= \int dx^2 \left[- \frac{1}{2} A_{0}{^{IJ}} \epsilon^{0ab} F_{abIJ} +\frac{A_{0}{^{IJ}}}{2} [e_{Ia} \Pi{_{J}}^{a}- e_{Ja} \Pi{_{I}}^{a} ] - 2e{^I}_{0}D_a\Pi{_{I}}^{a}  \right],
\label{eq27}
\end{equation}
and the primary Hamiltonian is given as
\begin{equation}
H_P= H_c + \int dx^2 \left[  \lambda^{I}{_{\alpha}} \phi{_{I}}^{\alpha} + \lambda^{IJ}{_{\alpha}} \phi{_{IJ}}^{\alpha}  \right],
\label{eq28}
\end{equation}
where $ \lambda^{I}{_{\alpha}},  \lambda^{IJ}{_{\alpha}}$ are Lagrange multipliers enforcing the constraints. For this field theory,  the non-vanishing fundamental Poisson brackets are
\begin{eqnarray}
\{ e_{\alpha} {^{I}}(x),\Pi^{\beta}{_{J}}(y) \}  &=& \delta {^ \beta} _\alpha \delta {^ I} _J \delta^2(x-y), \nonumber \\
\{ A_{\alpha} {^{IJ}}(x),\Pi^{\beta}{_{KL}}(y) \} &=& \frac{1}{2} \delta {^ \beta} _\alpha \left( \delta^I {_K} \delta^J{_L} - \delta^I{ _L} \delta^J{_K } \right) \delta^2(x-y).
\label{eq29}
\end{eqnarray}
The 18$\times$18 matrix whose entries are the Poisson brackets among the constraints (\ref{eq26a})
\begin{eqnarray}
\{\phi{_{I}}^{a} (x),\phi{_{J}}^{b} (y) \} &=& -\Lambda \epsilon^{0ab} \eta_{IJ}\delta^2(x-y), \nonumber \\
\{\phi{_{IJ}}^{a} (x),\phi{_{KL}}^{b} (y) \} &=& \frac{1}{2}\epsilon^{0ab}\left(\eta_{IL}\eta_{JK}- \eta_{IK}\eta_{JL} \right )\delta^2(x-y),
\label{eq30}
\end{eqnarray}
which  has rank=12 and 6 null-vectors. By using the 6 null-vectors and  consistency conditions,  one obtains   the following 6 secondary constraints
\begin{eqnarray}
\gamma^{0}{_{I}}&=&   \Pi{_{I}}^{0} \approx 0,  \nonumber \\
\gamma^{0}{_{IJ}}&=&   \Pi{_{IJ}}^{0}  \approx 0,  \nonumber \\
\dot{\phi}^{0}{_{IJ}}&=& \{\phi{_{IJ}}^{0} (x), {H}_{P} \} \approx 0 \quad \Rightarrow \quad \psi_{IJ}:= \frac{1}{2} [\epsilon^{0ab}F_{IJab}+ e_{Ja} \Pi{_{I}}^{a}- e_{Ia} \Pi{_{J}}^{a} ]   \approx 0, \nonumber \\
\dot{\phi}^{0}{_{I}}&=& \{\phi{_{I}}^{0} (x), {H}_{P} \} \approx 0 \quad \Rightarrow \quad \psi_{I}:= 2D_a\Pi{_{I}}^{a}  \approx 0,
\label{eq31}
\end{eqnarray}
and the rank allows us to fix the following  values for the Lagrangian multipliers
\begin{eqnarray}
\dot{\phi}{_{I}}^{a}&=&\{\phi{_{I}}^{a},H_{P}\}\approx0 \Rightarrow -\Lambda\epsilon^{0ab}(\lambda{_{b}}^{I}+D_{b}e{_{0}}^{I})\approx0, \nonumber \\
\dot{\phi}{_{IJ}}^{a}&=&\{\phi{_{IJ}}^{a},H_{P}\}\approx0 \Rightarrow \epsilon^{0ab}(\lambda{_{b}}^{IJ}-D_{b}A{_{0}}^{IJ})\approx0.
\label{eq32}
\end{eqnarray}

Consistency requires that their conservation in  time vanishes as well. For this theory there are no, third constraints. At this point,  we need to identify from  the primary and secondary constraints which one corresponds to the first and the second class. For this aim,  we need to calculate the rank and the null-vectors of the  24$\times$ 24 matrix whose entries will be the Poisson brackets between primary and secondary constraints, the non-zero brackets are given by
\begin{eqnarray}
\{\phi{_{I}}^{a} (x),\phi{_{J}}^{b} (y) \} &=& -\Lambda\epsilon^{0ab}\eta_{IJ}\delta^{2}(x-y), \nonumber \\
\{\phi{_{I}}^{a} (x),\psi_{J} (y) \} &=& -\Lambda\epsilon^{0ab}\left[\eta_{IJ}\partial_{b}\delta^{2}(x-y)-A_{bIJ}\delta^{2}(x-y)\right],\nonumber \\
\{\phi{_{IJ}}^{a} (x),\phi{_{KL}}^{b} (y) \} &=&  \frac{1}{2}\epsilon^{0ab}\left[ \eta_{IL}\eta_{JK}-\eta_{IK}\eta_{JL} \right] \delta^2(x-y) , \nonumber \\
\{\phi{_{IJ}}^{a} (x), \psi_{K}(y) \} &=& \left[ \Pi_{I}^{a}\eta_{JK}-\Pi_{J}^{a}\eta_{IK}\right]\delta^2(x-y) , \nonumber \\
\{\phi{_{IJ}}^{a} (x), \psi_{KL}(y) \} &=& \frac{\epsilon^{0ac}}{2} [A_{cIL}\eta_{JK}-A_{cJL}\eta_{IK}-A_{cKJ}\eta_{IL}+A_{cKI}\eta_{JL}\nonumber\\ &\,&+(\eta_{IK}\eta_{JL}-\eta_{IL}\eta_{JK})\partial_{c}]\delta^2(x-y),\nonumber\\
\{ \psi_{I} (x), \psi_{KL}(y) \} &=& \partial_{a}\delta^{2}(x-y) \left[\eta_{IK}\Pi_{L}^{a}-\eta_{IL}\Pi_{K}^{a}\right]+\delta^{2}(x-y)\left[A_{aLI}\Pi_{K}^{a}-A_{aKI}\Pi_{L}^{a}\right],\nonumber \\
\{ \psi_{IJ} (x), \psi_{KL}(y) \} &=&
\frac{1}{4}\big[\eta_{IK}(\Pi_{L}^{a}e_{Ja}-\Pi_{J}^{a}e_{La})+\eta_{JL} (\Pi_{K}^{a}e_{Ia}-\Pi_{I}^{a}e_{Ka})+\eta_{KJ}(\Pi_{I}^{a}e_{La}-\Pi_{L}^{a}e_{Ia})\nonumber\\ &\,& + \eta_{IL}(\Pi_{J}^{a}e_{Ka}-\Pi_{K}^{a}e_{Ja})\big]\delta^{2}(x-y).\nonumber
\label{eq33}
\end{eqnarray}
This matrix has rank=12 and 12 null vectors, thus, we find that our theory presents a set of 12 first class constraints and 12 second class constraints. By using the contraction of the null vectors with the constraints (\ref{eq26a}) and (\ref{eq31}), we identified the following 12 first class constraints
\begin{eqnarray}
\gamma^{0}{_{I}}&=&   \Pi{_{I}}^{0} \approx 0,  \nonumber \\
\gamma^{0}{_{IJ}}&=&   \Pi{_{IJ}}^{0}  \approx 0,  \nonumber \\
\gamma_{I}&=&-2 D_a\Pi{_I}^a + D_a  \phi^a{_{I}} + \Lambda e^J{_a}\phi _{IJ}{^a}, \nonumber \\
\gamma_{IJ}&=& D_a \phi^a{_{IJ}} + \frac{\epsilon^{0ab}}{2}F_{IJab} + \frac{1}{2}[\Pi{_I}^ae_{Ja}- \Pi{_J}^ae_{Ia} ],
\label{eq3b}
\end{eqnarray}
and the following 12 second class constraints
\begin{eqnarray}
\chi{_{I}}^{a}&=&   \Pi{_{I}}^{a}- \frac{\Lambda}{2}\epsilon^{0ab}e_{Ib} \approx 0,  \nonumber \\
\chi{_{IJ}}^{a}&=&   \Pi{_{IJ}}^{a}- \frac{\epsilon^{0ab}}{2}A_{bIJ}   \approx 0,
\label{eq4b}
\end{eqnarray}
It is important to remark  that these constraints have not been reported in the literature, and its complete structure defined on the full phase space  will be relevant in order to know the fundamental gauge transformations. On the other hand, the constraints will play a key role  to make progress in the quantization. All this information is only possible  by performing a pure Dirac's analysis.  \\
Now, we will calculate the algebra of the constraints
\begin{eqnarray}
\{\chi{_{I}}^{a} (x),\chi{_{J}}^{b} (y) \} &=& -\Lambda\epsilon^{0ab}\eta_{IJ}\delta^{2}(x-y), \nonumber \\
\{\chi{_{I}}^{a} (x),\gamma_{J} (y) \} &=& \Lambda\chi{_{IJ}}^{a}\delta^{2}(x-y)\approx0,\nonumber \\
\{\chi{_{I}}^{a} (x),\gamma_{JN} (y) \} &=& \frac{1}{2}\left[\eta_{IJ}\chi{_{N}}^{a}-\eta_{IN}\chi{_{J}}^{a}\right]\delta^{2}(x-y)\approx0, \nonumber \\
\{\chi{_{IJ}}^{a} (x), \gamma_{L}(y) \} &=& \frac{1}{2}\left[\eta_{IL}\chi_{J}^{a}-\eta_{JL}\chi_{I}^{a}\right]\delta^{2}(x-y)\approx0, \nonumber \\
\{\chi{_{IJ}}^{a} (x), \gamma_{KL}(y) \} &=&\frac{1}{2}\left[\chi{_{IL}^{a}}\eta_{KJ}-\chi{_{JL}^{a}}\eta_{KI}+\chi{_{KI}^{a}}\eta_{LJ}-\chi{_{KJ}^{a}}\eta_{LI}\right] \delta^2(x-y)\approx0 , \nonumber \\
\{\chi{_{IJ}}^{a} (x), \chi{_{KL}}^{b}(y) \} &=&\frac{1}{2}\epsilon^{0ab}\left[ \eta_{IL}\eta_{JK}-\eta_{IK}\eta_{JL} \right] \delta^2(x-y) , \nonumber \\
\{ \gamma_{I} (x), \gamma_{J}(y) \} &=&  \Lambda\gamma_{IJ}\delta^{2}(x-y)\approx0, \label{eq17c}  \\
\{ \gamma_{I} (x), \gamma_{KL}(y) \} &=& \frac{1}{2}\left[\gamma_{K}\eta_{IL}-\gamma_{L}\eta_{IK}\right]\delta^{2}(x-y)\approx0, \\
\{ \gamma_{IJ} (x), \gamma_{KL}(y) \} &=& -\frac{1}{2}\left[\gamma_{IL}\eta_{JK}-\gamma_{IK}\eta_{JL}+\gamma_{JK}\eta_{IL}-\gamma_{JL}\eta_{IK}\right]\delta^2(x-y)\approx0.
\label{eq36}
\end{eqnarray}
Where we are able to appreciate  that the algebra of the  first class constraints is closed and we do not need  conditions on the $\epsilon^{IJK}$ in order to obtain that algebra, this result is different from general relativity expressed by means of Palatini's theory,  because in Palatini's theory in order to obtain a closed algebra it is necessary to use the fact  $\epsilon^{IJK}$ are the structural constants of $SO(2,1)$ \cite{14}. Moreover,  because of (\ref{eq17c}) the algebra  does not form an $ISO(2,1)$ Poincar\'e algebra, however, it is a Lie algebra. In this respect, we are able to observe that Palatini's gravity without a cosmological constant forms a $ISO(2,1)$ Poincar\'e algebra \cite{13, 14};  in the exotic action, the cosmological constant can not be zero, this  will be  seen  in the following lines.   \\
We have developed a pure Dirac's analysis and there are  second class constraints,  however, they can  be eliminated through  Dirac's   bracket for the theory. In fact,  by obser\-ving that the matrix whose elements are only the Poisson brackets among second class constraints is given by
\begin{eqnarray}
C_{\alpha\beta}=
\left(
 \begin{array}{cccc}
   0& -\Lambda\eta_{IJ}&0&0\\
   \Lambda\eta_{IJ}&0&0&0\\
   0& 0&0& \frac{1}{2}\left[\eta_{IL}\eta_{JK}-\eta_{IK}\eta_{JL}\right]\\
   0& 0& -\frac{1}{2}\left[\eta_{IL}\eta_{JK}-\eta_{IK}\eta_{JL}\right]&0\\
  \end{array}
\right)\epsilon^{0ab}\delta^2(x-y),
\end{eqnarray}
its inverse will be
\begin{eqnarray}
C{^{-1}}_{\alpha\beta}=
\left(
 \begin{array}{cccc}
   0& \frac{1}{\Lambda}\eta^{IJ}&0&0\\
   -\frac{1}{\Lambda}\eta^{IJ}&0&0&0\\
   0& 0&0&-2\left[\eta^{IL}\eta^{JK}-\eta^{IK}\eta^{JL}\right]\\
   0& 0& 2\left[\eta^{IL}\eta^{JK}-\eta^{IK}\eta^{JL}\right]&0\\
  \end{array}
\right)\epsilon{_{0ab}}\delta^2(x-y).
\end{eqnarray}
The Dirac's  brackets among two functionals $A$, $B$ are expressed by
\begin{eqnarray}
\{A(x),B(y)\}_{D}=\{A(x),B(y)\}_{P}+\int dudv\{A(x),\zeta^{\alpha}(u)\}C{^{-1}}_{\alpha\beta}(u,v)\{\zeta^{\beta}(v),B(y)\},
\end{eqnarray}
where $\{A(x),B(y)\}_{P}$ is the usual Poisson brackets between the functionals $A$, $B$ and   $\zeta^{\alpha}(u)=(\chi{_{I}}^{a},\chi{_{IJ}}^{a})$. Hence, we obtain the following Dirac's brackets of the theory
\begin{eqnarray}
\{e{^{I}}_{a}(x),\Pi{^{b}}_{J}(y)\}_{D}&=&\{e{^{I}}_{a}(x),\Pi{^{b}}_{J}(y)\}_{P}+\int dudv\{e{^{I}}_{a}(x),\zeta^{\alpha}(u)\}C{^{-1}}_{\alpha\beta}(u,v)\{\zeta^{\beta}(v),\Pi{^{b}}_{J}(y)\},  \nonumber \\
&=&\frac{1}{2}\delta{^{b}}_{a}\delta{^{I}}_{J}\delta^{2}(x-y),
\label{eq39}
\end{eqnarray}

\begin{eqnarray}
\{e{^{I}}_{a}(x),e{^{J}}_{b}(y)\}_{D}&=&\{e{^{I}}_{a}(x),e{^{J}}_{b}(y)\}_{P}+\int dudv\{e{^{I}}_{a}(x),\zeta^{\alpha}(u)\}C{^{-1}}_{\alpha\beta}(u,v)\{\zeta^{\beta}(v),e{^{J}}_{b}(y)\},  \nonumber \\
&=&\frac{1}{\Lambda}\eta{^{IJ}}\epsilon_{0ab}\delta^{2}(x-y),
\label{eq40}
\end{eqnarray}

\begin{eqnarray}
\{\Pi{^{a}}_{I}(x),\Pi{^{b}}_{J}(y)\}_{D}&=&\{\Pi{^{a}}_{I}(x),\Pi{^{b}}_{J}(y)\}_{P}+\int dudv\{\Pi{^{a}}_{I}(x),\zeta^{\alpha}(u)\}C{^{-1}}_{\alpha\beta}(u,v)\{\zeta^{\beta}(v),\Pi{^{b}}_{J}(y)\},  \nonumber \\
&=&\frac{\Lambda}{4}\eta_{IJ}\epsilon{^{0ab}}\delta^{2}(x-y),
\label{eq41}
\end{eqnarray}

\begin{eqnarray}
\{A{^{IJ}}_{a}(x),\Pi{^{b}}_{LN}(y)\}_{D}&=&\{A{^{IJ}}_{a}(x),\Pi{^{b}}_{LN}(y)\}_{P}+\int dudv\{A{^{IJ}}_{a}(x),\zeta^{\alpha}(u)\}C{^{-1}}_{\alpha\beta}(u,v)\{\zeta^{\beta}(v),\Pi{^{b}}_{LN}(y)\},  \nonumber \\
&=&\frac{1}{4}\delta{^{b}}_{a}\left[\delta{^{I}}_{L}\delta{^{J}}_{N}-\delta{^{I}}_{N}\delta{^{J}}_{L} \right]\delta^{2}(x-y),
\label{eq42}
\end{eqnarray}

\begin{eqnarray}
\{A{^{IJ}}_{a}(x),A{^{LN}}_{b}(y)\}_{D}&=&\{A{^{IJ}}_{a}(x),A{^{LN}}_{b}(y)\}_{P}+\int dudv\{A{^{IJ}}_{a}(x),\zeta^{\alpha}(u)\}C{^{-1}}_{\alpha\beta}(u,v)\{\zeta^{\beta}(v),A{^{LN}}_{b}(y)\},  \nonumber \\
&=&\frac{1}{2}\left[\eta{^{IL}}\eta{^{JN}}-\eta{^{IN}}\eta{^{JL}}\right]\epsilon{_{0ab}}\delta^{2}(x-y),
\label{eq43}
\end{eqnarray}

\begin{eqnarray}
\{\Pi{^{a}}_{IJ}(x),\Pi{^{b}}_{LN}(y)\}_{D}&=&\{\Pi{^{a}}_{IJ}(x),\Pi{^{b}}_{LN}(y)\}_{P}+\int dudv\{\Pi{^{a}}_{IJ}(x),\zeta^{\alpha}(u)\}C{^{-1}}_{\alpha\beta}(u,v)\{\zeta^{\beta}(v),\Pi{^{b}}_{LN}(y)\},  \nonumber \\
&=&\frac{1}{8}\left[\eta_{{IL}}\eta_{{JN}}-\eta_{{IN}}\eta_{{JL}}\right]\epsilon{^{0ab}}\delta^{2}(x-y).
\label{eq44}
\end{eqnarray}

\begin{eqnarray}
\{e{^{I}}_{a}(x),A{^{LN}}_{b}(y)\}_{D}&=&\{e{^{I}}_{a}(x),,A{^{LN}}_{b}(y)\}_{P}+\int dudv\{e{^{I}}_{a}(x),\zeta^{\alpha}(u)\}C{^{-1}}_{\alpha\beta}(u,v)\{\zeta^{\beta}(v),,A{^{LN}}_{b}(y)\},  \nonumber \\
&=&0,
\end{eqnarray}

\begin{eqnarray}
\{e{^{I}}_{a}(x),\Pi{^{b}}_{LN}(y)\}_{D}&=&\{e{^{I}}_{a}(x),\Pi{^{b}}_{LN}(y))\}_{P}+\int dudv\{e{^{I}}_{a}(x),\zeta^{\alpha}(u)\}C{^{-1}}_{\alpha\beta}(u,v)\{\zeta^{\beta}(v),\Pi{^{b}}_{LN}(y)\},  \nonumber \\
&=&0,
\end{eqnarray}

\begin{eqnarray}
\{A{^{IJ}}_{a}(x),\Pi{^{b}}_{L}(y)\}_{D}&=&\{A{^{IJ}}_{a}(x),\Pi{^{b}}_{L}(y))\}_{P}+\int dudv\{A{^{IJ}}_{a}(x),\zeta^{\alpha}(u)\}C{^{-1}}_{\alpha\beta}(u,v)\{\zeta^{\beta}(v),\Pi{^{b}}_{L}(y)\},  \nonumber \\
&=&0,
\end{eqnarray}
It is important to remark that the fields $e$,  $A$ and their canonical momenta   have become non-commutative, and the cosmological constant can not be fixed to  zero. On the other hand, in Palatini's  gravity, by performing a pure Hamiltonian analysis, Dirac's brackets among the fields $e$ and $A$ become  commutative and the cosmological constant can  be taken as zero (see \cite{14}). This result marks  a difference at the classical level among exotic  and Palatini  actions. Furthermore, we notice that  the term of Nieh-Yang becomes a magnetic like term,  just as is present in Landau's problem. In fact,  for a charged particle of mass $m$  confined by a quadratic potential that  moves   in a uniform  magnetic field,  the Lagrangian is given by \cite{15}
\begin{equation}
L= \frac{m}{2} x_i^2 + \frac{B}{2} \epsilon_{ij} \dot{x}^i x^j- \frac{K}{2} x_i^2,
\label{eq30a}
\end{equation}
where $B$ is the magnetic field and $K$ is a constant. In general the action (\ref{eq30a}) is not singular and the Hamiltonian analysis is easy  to carry out.  Because of the action is not singular, we can take $B$ or $K$ as zero without problem. However,   by taking the limit $m\rightarrow0$ the system becomes  singular and after a Dirac's analysis of (\ref{eq30a}) in that limit, there are second class constraints, and  the coordinates  are  non-commutative;  Dirac's brackets  of the theory  are given by  $\{ x^ i, x^ j  \}_D = - \frac{\epsilon^{ij}}{B}$, thus, for this singular theory  $B$ can not be zero;  in fact, the spectra of energy depend on a factor $\frac{1}{B}$ \cite{12}.   In this manner, in analogy with the action (\ref{eq30a}),  in the  exotic action  the Nieh-Yang term  is a  "magnetic field"  like  term (see the term  $ \frac{\Lambda}{2} \epsilon^{0ab} \dot{e}^I_a e_{Ib}$ of (\ref{eq22})), namely, the cosmological constant becomes to be the 	 magnetic field $B$ and the field $e$ the non-commutative coordinates (see eq. (\ref{eq40})). Of course,  the Chern-Simons term can be treated in the same form;  however,  Chern-Simons gives non-commutative connections $A$ (see eq (\ref{eq43})), and the Nieh-Yang term gives non-commutative fields $e$. Thus, the Nieh-Yang term becomes  a non-commutative gauge theory for the triad field.  Therefore,    we realize  that for a singular theory it  is not a correct step  to  fix  the parameters  that occur   in the theory    before developing  a detailed analysis.  In order to study a singular system with arbitrary parameters,  first,   it is mandatory to perform a detailed Dirac's analysis,  then, we could study the behavior of the action  by taking the limit $\rightarrow 0$ of the parameters. The exotic action is a singular system and our detailed analysis indicates that the cosmological constant cannot be fixed to  zero.    \\
Moreover, the identification of the constraints will allow us to identify the extended action. By using the first class constraints $(17)$, the second class constraints $(18)$, and the Lagrangian multipliers $(15)$ we find that the extended action takes the form
\begin{eqnarray}
S_{E}[e{^{I}}_{\alpha},A{^{IJ}}_{\alpha},\Pi{^{\alpha}}{_{I}},\Pi^{\alpha}{_{IJ}},u{_{0}}^{I},u{_{0}}^{IJ},u^{I},u^{IJ},v{_{a}}^{I},v{_{a}}^{IJ}]&=&\int_{M}\big[ \dot{e}{^{I}}_{\alpha}\Pi{^{\alpha}}{_{I}}+\dot{A}{^{IJ}}_{\alpha}\Pi^{\alpha}{_{IJ}}\nonumber\\-H'-u{_{0}}^{I}\gamma{_{I}}^{0}-u{_{0}}^{IJ}\gamma{_{IJ}}^{0}-u^{I}\gamma{_{I}}-u^{IJ}\gamma{_{IJ}}-v{_{a}}^{I}\chi{_{I}}^{a}-v{_{a}}^{IJ}\chi{_{IJ}}^{a}\big]dx^{3},
\end{eqnarray}
where $H^{'}$ is the linear combination of first class constraints
\begin{eqnarray}
H^{'}=\int\left[e_{0}{^{I}}\gamma{_{I}}-A_{0}{^{IJ}}\gamma{_{IJ}}\right]dx^{2},
\end{eqnarray}
and $u{_{0}}^{I},u{_{0}}^{IJ},u^{I},u^{IJ},v{_{a}}^{I},v{_{a}}^{IJ}$ are Lagrange multipliers enforcing the first and second class constraints.
From the extended action we can identify the extended Hamiltonian given by
\begin{eqnarray}
H_{E}=H^{'}+\int\left[u{_{0}}^{I}\gamma^{0}{_{I}}+u{_{0}}^{IJ}\gamma^{0}{_{IJ}}+u^{I}\gamma{_{I}}+u^{IJ}\gamma{_{IJ}}\right]dx^{2}.
\end{eqnarray}
It is important to remark, that the theory under study has an extended Hamiltonian which is a linear combination of first class constraints reflecting the general covariance of the theory, just as General Relativity, thus, in order to perform a quantization of  the theory,  it is not possible to construct the Schr\"{o}dinger equation because the
action of the Hamiltonian on physical states is annihilation. In Dirac's quantization of systems with general covariance, the restriction of our physical state is archived by demanding that the first class constraints in their quantum form must be satisfied and the Dirac's brackets must be taken into account as well, thus in this paper we have all the tools for  studying  the quantization of the theory by means of a canonical framework. \\
One of the most important symmetries that  can be studied  by using the Hamiltonian method, are the gauge transformations. Gauge transformations are fundamental in the identification of  physical observables \cite{5}. In this respect, we have commented above that a detailed analysis will give us the correct gauge symmetry. In fact, the correct gauge symmetry is  obtained  according to Dirac's conjecture  by constructing a gauge generator  using the first class constraints, and the structure of the constraints  defined on the full phase space   will give us the fundamental gauge transformations.  For this aim, we will apply the Castellani's algorithm  to construct the gauge generator.
We define the generator of gauge transformations as
\begin{eqnarray}
G=\int_{\sum}\left[D{_{0}}\varepsilon{^{I}}_{0}\gamma{^{0}}_{I}+D{_{0}}\varepsilon{_{0}}^{IJ}\gamma{^{0}}_{IJ}+\varepsilon^{I}\gamma{_{I}}+\varepsilon^{IJ}\gamma{_{IJ}}\right].
\end{eqnarray}
Therefore, we find that the gauge transformations on the phase space are
\begin{eqnarray}
\delta{_{0}} e{^{I}}_{0}&=&D_{0}\varepsilon{^{I}}_{0},\nonumber \\
\delta{_{0}} e{^{I}}_{a}&=&D{_{a}}\varepsilon^{I}+\varepsilon^{IJ}e{_{aJ}},\nonumber\\
\delta{_{0}} A_{0}{^{IJ}}&=&D_{0}\varepsilon{_{0}}{^{IJ}},\nonumber\\
\delta{_{0}} A_{a}{^{IJ}}&=&\frac{\Lambda}{2}\left[e{^{J}}_{a}\varepsilon^{I}-e{^{I}}_{a}\varepsilon^{J}\right]-D{_{a}}\varepsilon^{IJ},\nonumber\\
\delta{_{0}} \Pi{^{0}}_{I}&=& 0,\nonumber \\
\delta{_{0}} \Pi{^{a}}_{I}&=& \frac{\Lambda}{2}\epsilon^{0ab}\partial{_{b}}\varepsilon_{I}+\Lambda\varepsilon^{J}\Pi{^{a}}_{IJ}-\varepsilon{^{J}}{_{I}}\Pi^{a}_{J},\nonumber\\
\delta{_{0}} \Pi{^{0}}_{IJ}&=&-\varepsilon{_{I}}^{L}\Pi{^{0}}_{LJ}+\varepsilon{_{J}}^{L}\Pi{^{0}}_{LI},\nonumber\\
\delta{_{0}}
\Pi{^{a}}_{IJ}&=&\frac{1}{2}\left[\varepsilon{_{I}}\Pi{^{a}}{_{J}}-\varepsilon{_{J}}\Pi{^{a}}{_{I}}\right]+\left[\varepsilon{_{J}}^{L}\Pi{^{a}_{IL}}-\varepsilon{_{I}}^{L}\Pi{^{a}_{JL}}\right]+\frac{1}{2}\epsilon^{0ba}\partial_{b}\varepsilon_{IJ}.
\label{eq50}
\end{eqnarray}
We realize  that the fundamental  gauge transformations of the exotic action are given by (\ref{eq50}) and do not correspond to diffeomorphisms,  but they are $\Lambda$-deformed $ISO(2,1)$ Poincar\'e transformations.  However, any theory with  a dynamical background metric is diffeomorphisms covariant, and this symmetry must be obtained from the fundamental gauge transformation. Hence, the   diffeomorphisms can be found  by  redefining the gauge parameters as $\varepsilon{_{0}}^{I}=\varepsilon^{I}=\xi^{\rho}e{^{I}}_{\rho}$, $\varepsilon{_{0}}^{IJ}=\varepsilon^{IJ}=-\xi^{\rho}A_{\rho}{^{IJ}}$,  and  the gauge transformation (\ref{eq50}) takes the following form
\begin{eqnarray}
e'{^{I}}{_{\alpha}}&\rightarrow&e{^{I}}{_{\alpha}}+\mathfrak{L}{_{\xi}}e{^{I}}{_{\alpha}}+\xi^{\rho}\left[D{_{\alpha}}e{^{I}}{_{\rho}}-D{_{\rho}}e{^{I}}{_{\alpha}}\right] ,  \nonumber \\
A'{_{\alpha}}^{IJ}&\rightarrow&A{_{\alpha}}^{IJ}+\mathfrak{L}{_{\xi}}A{_{\alpha}}^{IJ}+\xi^{\rho}\left[R{^{IJ}}{_{\alpha\rho}}-\frac{\Lambda}{2}(e{^{I}}_{\alpha}e{^{J}}_{\rho}-e{^{J}}_{\alpha}e{^{I}}_{\rho})\right] ,
\label{eq51}
\end{eqnarray}
Therefore, diffeomorphisms are obtained (on shell)  from  the fundamental gauge transformations as  an internal symmetry of the theory. With the correct identification of the constraints, we can carry out the counting of degrees of freedom  in the following form: there are  36 canonical variables $(e{^I}_{\alpha}, A_{\alpha} {^{IJ}}, \Pi^{\alpha}{_{I}},\Pi^{\alpha}{_{IJ}})$, $12$ first class constraints $(\gamma{_{I}}^{0}, \gamma{_{IJ}}^{0}, \gamma_{I}, \gamma_{IJ})$ and $12$ second class constraints $(\chi{_{I}}^{a}, \chi{_{IJ}}^{a})$  and  one  concludes that the  exotic action for gravity  in three   dimensions is devoid of degrees of freedom, therefore, the theory is topological.\\
As a conclusion of this part, we have performed  a pure  Hamiltonian analysis for the exotic action by working with the complete configuration space. With the present analysis, we have obtained the extended action, the extended Hamiltonian, the complete structure of the constraints  on the full phase space, and the algebra among them. The price to pay for working on the complete phase space,  is that  the  theory presents a set of first and second class constraints;  by using the second class constraints we have constructed  Dirac's brackets and they will be useful in the quantization of the theory.
\section{The symplectic method  for Exotic action }
In order to develop a complete analysis, in this section we shall  carry out  the covariant canonical formalism for the theory, and we shall  confirm some results obtained in the above section.  \\
Let us start by  calculating the variation of the  exotic action
 \begin{eqnarray}
\delta S[A,e]_{exotic}&=& \int_M\left[ \frac{1}{2}  \left( \epsilon^{\alpha \mu\nu} F_{IJ \mu\nu}[A] - \Lambda \epsilon^{\alpha \mu\nu}e_{I \mu} e_{J \nu} \right) \delta A_{\alpha}{^ {IJ}}  + \left( \Lambda  \epsilon^{\alpha \mu\nu} D_{\mu} e^I{_{\nu}} \right)\delta e{^I}_{ \alpha}   \right] \nonumber  \\
&-& \int_M \partial_\mu \left(\Lambda  \epsilon^{\mu \alpha \nu} e_{I \alpha} \delta e^I{_\nu} +   \epsilon^{\mu \alpha \nu}  A_{\alpha}{^{ IJ}} \delta A_{\nu IJ} \right),
\label{eq65}
\end{eqnarray}
where we can identify the equations of motion   (\ref{eq4}) and the integral kernel for  the symplectic structure from the boundary term
\begin{equation}
\Psi^\mu_{exotic}=  \epsilon^{\mu \alpha \nu} \Lambda e_{I \alpha}\delta e^I{_\nu} +   \epsilon^{\mu \alpha \nu}  A_{\alpha}{^{ IJ}} \delta A_{\nu IJ}.
\label{eq66}
\end{equation}
which does not contribute locally to the dynamics, but generates the symplectic form on the phase space \cite{4}.\\
Now, we define the fundamental concept in the studio
of the symplectic  formalism of the theory: the covariant phase space for the theory
described by (5) is the space of solutions of (\ref{eq4}), and we shall  call it $Z$. In this manner, we  can obtain the fundamental two-form  of the geometric structure for the theory
by means of the variation (exterior derivative on $Z$ \cite{4} ) of the symplectic potential (\ref{eq66})
\begin{equation}
 \varpi= \int_\Sigma  J^\mu d \Sigma_\mu= \int_\Sigma \delta \Psi^\mu d \Sigma_\mu =\int_\Sigma \left(\epsilon^{\mu \alpha \nu}\Lambda  \delta e_{I \alpha}\wedge \delta e^I{_\nu} +   \epsilon^{\mu \alpha \nu}  \delta A_{\alpha}{^{ IJ}} \wedge  \delta A_{\nu IJ} \right)d\Sigma_{\mu},
 \label{eq67}
 \end{equation}
 where $\Sigma$ is a Cauchy hypersurface. We are able to observe in the geometric structure  (\ref{eq67})  the non-commutative character of the dynamical variables.\\
 So, we  will prove that our symplectic form is closed and gauge invariant. Moreover,
the integral kernel of the geometric form $J^\mu$ is conserved, which guarantees
that $\varpi$ is independent of $\Sigma$. We need to remember that the closeness of $\varpi$   is equivalent  to the Jacobi identity that the Poisson brackets satisfy in the
usual Hamiltonian scheme. \\
 In order to prove  the closeness of $\varpi$ , we can observe that $\delta^2 e_\mu ^I=0$ and $\delta^2 A_{\alpha}{^{ IJ}} =0$,    because $ e_\mu ^I$ and $ A_{\alpha}{^{ IJ}} $ are independent 0-forms on the covariant phase space $Z$
and $\delta$ is nilpotent, thus
\begin{eqnarray}
 \delta \varpi &=& \int_\Sigma \bigg\{ \epsilon^{\mu \alpha \nu}\Lambda  \delta^2 e_{I \alpha}\wedge \delta e^I{_\nu} - \epsilon^{\mu \alpha \nu}\Lambda  \delta e_{I \alpha}\wedge \delta^2 e^I{_\nu}+   \epsilon^{\mu \alpha \nu}  \delta^2 A_{\alpha}{^{ IJ}} \wedge  \delta A_{\nu IJ} \nonumber \\ &-& \epsilon^{\mu \alpha \nu}  \delta A_{\alpha}{^{ IJ}} \wedge  \delta^2 A_{\nu IJ} \bigg\} d\Sigma_{\mu}=0,
 \label{eqa45}
 \end{eqnarray}
therefore the geometric form is closed. \\
For future useful calculations  we shall obtain the linearized equations of motion of  the theory. For this purpose, we replace $ A_{\nu IJ} \rightarrow \delta A_{\nu IJ}  $ and $e_\mu ^I  \rightarrow \delta e_\mu ^I  $ in (\ref{eq4}) and keep
only the first-order terms, we obtain
\begin{eqnarray}
\epsilon^{\mu \alpha \nu} \left[ D_{[\alpha}\delta A_{\nu] IJ} \right] -  \Lambda \left[ e_{\alpha I}\delta e_{\nu J} + \delta e_{\alpha I} e_{\nu J} \right]&=&0, \nonumber \\
\Lambda  \epsilon^{\mu \alpha \nu}  \left[ D_\alpha \delta e_{\nu I}  + e_{\nu }^J \delta A_\alpha {_{IJ}} \right]=0.
\label{eq46}
\end{eqnarray}
Furthermore, we can see that under  fundamental  gauge transformations given in (\ref{eq50}) and for some infinitesimal variation we have
\begin{eqnarray}
\delta A{'} _\alpha {^{IJ}} &=& \delta A_\alpha {^{IJ}}- \delta A_\alpha {^{ I}}{_K} \epsilon ^{KJ} -\delta A_\alpha {^{ J}}{_K} \epsilon ^{IK}, \nonumber \\
\delta e {'} _\alpha ^{I} &=& \delta e  _\alpha ^{I} + \delta e_{\alpha J} \epsilon ^{IJ},
\label{eq47}
\end{eqnarray}
thus,  by using (\ref{eq47}),   we find that $\varpi$ transforms
\begin{eqnarray}
\varpi{'} &=& \int_\Sigma \left(\epsilon^{\mu \alpha \nu}\Lambda  \delta e{'}_{I \alpha}\wedge \delta e{'}^I{_\nu} +   \epsilon^{\mu \alpha \nu}  \delta A{'}_{\alpha}{^{ IJ}} \wedge  \delta A{'}_{\nu IJ} \right)d\Sigma_{\mu} \nonumber \\
&=& \varpi + \int _\Sigma \textrm{O} (\epsilon^2) d\sigma.
\end{eqnarray}
Therefore, $\varpi$ is a $SO(2,1)$ singlet. This result allows us to prove that
\begin{eqnarray}
\partial_\mu J^\mu&=& D_\mu J^\mu= \Lambda \epsilon^{\mu \alpha \nu} \left[ D_\mu \delta e_{ \alpha I} \wedge \delta e_{\nu}{^I } +  \delta  e_{ \alpha I} \wedge D_\mu  \delta e_{\nu}{^I } + D_\mu  \delta_{\alpha}{^IJ} \wedge \delta _{\nu IJ} +  \delta_{\alpha}{^IJ} \wedge D_\mu \delta _{\nu IJ}  \right] \nonumber \\
&=& - \Lambda   \epsilon^{\mu \alpha \nu} e_{\alpha}{^J} \delta A_{\mu IJ} \wedge \delta e_{\nu}{^ I} - \Lambda  \epsilon^{\mu \alpha \nu}e_{\nu}{^J} \delta e_{\alpha I} \wedge \delta A_{\mu IJ} + \frac{\Lambda}{2} \epsilon^{\mu \alpha \nu} e_{\alpha J} \delta e_{\mu I} \wedge \delta A_{\nu IJ} \nonumber \\
&+&\frac{\Lambda}{2} \epsilon^{\mu \alpha \nu}   e_{\mu I} \delta e_{\alpha J} \wedge \delta A_{\nu}^{ IJ}+  \frac{\Lambda}{2} \epsilon^{\mu \alpha \nu}   e_{\mu I} \delta A_{\alpha}{^ {IJ}} \wedge \delta e_{\nu J} +  \frac{\Lambda}{2} \epsilon^{\mu \alpha \nu}   e_{\nu J} \delta A_{\alpha}{^{IJ}} \wedge \delta e_{\mu I}=0,
\end{eqnarray}
where we have used the  linearized  equations given in (\ref{eq46}),  and the antisymmetry of 1-forms  $\delta e_{\nu}{^ I}$ and $\delta A_{\alpha}{^{ IJ}}$. Therefore, $\varpi$ is independent of $\Sigma$, thus performing a Lorentz transformation $\Sigma_t$ $\rightarrow$ $ \Sigma_t^{'}$ and $\varpi$ $\rightarrow$ $\varpi'$
\begin{equation}
\varpi'=\int_\Sigma \delta \Psi'^\alpha d\Sigma'_\alpha =\int_\Sigma \delta \Psi^\alpha d\Sigma_\alpha= \varpi.
\end{equation}
In this manner, with these results we have constructed a Lorentz and gauge invariant symplectic structure on the  phase
space and it is possible to formulate the Hamiltonian theory in a manifestly covariant way.\\
In order to reproduce the gauge transformations found in the Hamiltonian formalism by using now the symplectic method,  let us  consider that upon picking $\Sigma$ to be the standard initial value surface $t=0$,  hence equation  (\ref{eq67})  takes the standard form
\begin{equation}
\varpi= \int_\Sigma \left[  \delta \Pi^a{_{I}} \wedge \delta e_a {^{I}} +  \delta \Pi ^a{_{IJ}} \delta A{_{a}}^{IJ}\right],
\label{eq46a}
\end{equation}
where $ \Pi^a{_{I}} \equiv \Lambda \epsilon^{0ab} e_{Ib} $ and $  \Pi ^a{_{IJ}} \equiv  \epsilon^{0ab}A_{bIJ}$. \\
For two 0-forms $f$, $g$ defined on $Z$, the Hamiltonian vector field defined by the symplectic form (\ref{eq46a})
\begin{equation}
X_f \equiv \int_\Sigma \frac{\delta f}{ \delta  \Pi^a{_{I}}} \frac{\delta }{ \delta e_a {^{I}} } -  \frac{\delta f }{ \delta e_a {^{I}} }\frac{\delta }{ \delta  \Pi^a{_{I}}} + \frac{\delta f}{\delta  \Pi ^a{_{IJ}}} \frac{\delta }{ \delta A{_{a}}^{IJ} } - \frac{\delta f }{ \delta A{_{a}}^{IJ} }\frac{\delta } {\delta  \Pi ^a{_{IJ}}} ,
\end{equation}
and the Poisson bracket $\{f,g \}= - X_f (g)$ is given by
\begin{equation}
\{f,g \} \equiv \int_\Sigma  \frac{\delta f }{ \delta e_a {^{I}} }\frac{\delta g }{ \delta  \Pi^a{_{I}}}  - \frac{\delta f}{ \delta  \Pi^a{_{I}}} \frac{\delta g }{ \delta e_a {^{I}} }  + \frac{\delta f }{ \delta A{_{a}}^{IJ} }\frac{\delta g } {\delta  \Pi ^a{_{IJ}}} - \frac{\delta f}{\delta  \Pi ^a{_{IJ}}} \frac{\delta g }{ \delta A{_{a}}^{IJ} }.
\end{equation}
Furthermore,  smearing the constraints (\ref{eq3b})  with test fields we obtain
\begin{eqnarray}
\gamma_{I}\left [C^I \right] &=&\int _\Sigma C^I\left[-2 D_a\Pi{_I}^a + D_a  \phi^a{_{I}} + \Lambda e^J{_a}\phi _{IJ}{^a} \right], \nonumber \\
\gamma_{IJ} \left [C^{I J}\right]&=& \int_ \Sigma C^{IJ}\left[ D_a \phi^a{_{IJ}} + \frac{\epsilon^{0ab}}{2}F_{IJab} + \frac{1}{2}[\Pi{_I}^ae_{Ja}- \Pi{_J}^ae_{Ia} ] \right].
\label{eq54}
\end{eqnarray}
By inspection, the functional derivatives different from  zero are given by
\begin{eqnarray}
\frac{\delta \gamma_{I}\left[C^I \right] }{\delta e_a{^{I}}}= - \frac{\Lambda}{2} \epsilon^{0ab} D_b C_I - \Lambda C^J \phi^a {_{IJ}} , \quad \quad \quad  \frac{\delta \gamma_{I} \left[C^I \right] }{\delta \Pi{_{I}} ^a}= D_a C^I, \nonumber \\
\frac{\delta \gamma_{IJ}\left[C^{IJ} \right] }{\delta e_a{^{I}}} = -C_{I}{^{J}} \Pi ^a_J, \quad \quad \quad\quad \quad \quad \quad  \frac{\delta \gamma_{IJ} \left[C^{IJ} \right] }{\delta \Pi{_{I}} ^a}=  C^{IJ} e_{a J}, \nonumber  \\
\frac{\delta \gamma_{I}\left[C^I \right] }{\delta A_a{^{IJ}}}= - \frac{1}{2} \left[ C_I \Pi{_{{J}} ^a} -    C_J \Pi{_{I}} ^a \right], \quad \quad \quad  \frac{\delta \gamma_{I}\left[C^I \right] }{\delta \Pi ^a{_{IJ}}}= \frac{\Lambda}{2} \left[ C^I e_a ^J -C^J e_a ^I  \right] ,  \nonumber \\
\frac{\delta \gamma_{IJ}\left[C^{IJ} \right] }{\delta A_a{^{IJ}}}= \frac{\epsilon ^{0ab}}{2 } \partial_b C_{IJ} + \left[ C_{I} {^{F}} \Pi ^a_{JF} - C_{J} {^{F}} \Pi ^a_{IF} \right], \quad \quad \quad   \frac{\delta \gamma_{IJ}\left[C^{I J}\right] }{\delta \Pi ^a{_{IJ}}} = - D_a C^{IJ}.
\label{eq50a}
\end{eqnarray}
Thus, by using (\ref{eq54})  and  (\ref{eq50a}) (the motion on $Z$ generated by $ \gamma_{I}\left[C^I \right] $ and $ \gamma_{IJ}\left[C^{I J}\right]$, is given by
\begin{eqnarray}
e{^{I}}_{a}&\rightarrow& e{^{I}}_{a}+ \xi D{_{a}}C^{I}+ \xi C^{IJ}e{_{aJ}} + O(\xi^2),\nonumber\\
 A_{a}{^{IJ}}&\rightarrow&A_{a}{^{IJ}}-\frac{\xi \Lambda}{2}\left[e{^{I}}_{a}C^{J}-e{^{J}}_{a}C^{I}\right]-\xi D{_{a}}C^{IJ} + O(\xi^2),\nonumber\\
 \Pi{^{a}}_{I}&\rightarrow&\Pi{^{a}}_{I}+  \frac{\xi \Lambda}{2}\epsilon^{0ab}\partial{_{b}}C_{I}+\xi \Lambda C^{J}\Pi{^{a}}_{IJ}-\xi C{^{J}}{_{I}}\Pi^{a}_{J}+  O(\xi^2),,\nonumber\\
\Pi{^{a}}_{IJ}&\rightarrow&\Pi{^{a}}_{IJ} + \frac{\xi}{2}\left[C{_{I}}\Pi{^{a}}{_{J}}-C{_{J}}\Pi{^{a}}{_{I}}\right]+ \xi \left[C{_{J}}^{L}\Pi{^{a}_{IL}}-C{_{I}}^{L}\Pi{^{a}_{JL}}\right]+\frac{\xi}{2}\epsilon^{0ba}\partial_{b}C_{IJ}+  O(\xi^2),
\label{eq56a}
\end{eqnarray}
where $\xi$ is an infinitesimal parameter \cite{16}. We are able to observe that the gauge transformations (\ref{eq56a})  are  those found using Dirac's method (see (\ref{eq50})),  and correspond to  $\Lambda$-deformed  Poincar\'e  transformations. Furthermore, it is well-known that any background independent theory is diffeomorphisms covariant and this symmetry should be manifest in our geometric structure, in order to prove that  $\varpi$ is diffeomorphisms covariant  we observe that (\ref{eq51}) for some infinitesimal variation takes the form
\begin{eqnarray}
\delta e'{^{I}}{_{\alpha}}&\rightarrow&\delta e{^{I}}{_{\alpha}} + \xi^\mu \partial_\mu \delta  e{^{I}}_{\alpha} +   \delta  e{^{I}} _{\mu}\partial_\alpha \xi^\mu, \nonumber \\
\delta A'_{\alpha}{^{IJ}}&\rightarrow&\delta A_{\alpha}{^{IJ}} + \xi^\mu \partial_\mu \delta A_{\alpha}{^{IJ}} + \delta A_{\mu}{^{IJ}}  \partial_\alpha  \xi^\mu,
\label{eq57aa}
\end{eqnarray}
thus by using (\ref{eq57aa}),   $\varpi$ will undergo the transformation as
\begin{eqnarray}
\varpi' &=& \int_\Sigma \left(\epsilon^{\mu \alpha \nu}\Lambda  \delta e'_{I \alpha}\wedge \delta e'^I{_\nu} +   \epsilon^{\mu \alpha \nu}  \delta A'_{\alpha}{^{ IJ}} \wedge  \delta A'_{\nu IJ} \right)d\Sigma_{\mu}, \nonumber \\
&=& \varpi + \int_\Sigma \mathfrak{L}{_{\xi}}\varpi.
\end{eqnarray}
Moreover, $\mathfrak{L}{_{\xi}}\varpi = \xi  \cdot d\varpi + d(\xi \cdot  \varpi)$, but $\varpi$ is closed ($d\varpi=0$), hence the term  on the right hand side  is a surface term. Therefore, we have showed that $\varpi$ is invariant under infinitesimal diffeomorphisms.
As a conclusion of this section, we have constructed a gauge invariant
symplectic form on $Z$ which in turn represents a complete Hamiltonian description of
the covariant phase space for the theory,  and it will allow us to analyze the quantum treatment
in forthcoming works.\\
\section{ Conclusions and prospects}
In this paper, a detailed  Hamilton analysis for an exotic action has been performed; in our analysis,  the price to pay by working on the complete phase space is that the theory presents a set of first and second class constraints and  we have identified  their  full structure. By identifying the complete structure of the constraints, we found the fundamental gauge transformations of the theory  corresponding to deformed Poincar\'e  transformations and by defining the gauge parameters,  diffeomorphisms can be  obtained from the fundamental gauge symmetry. It is important to comment, that only by using a pure Dirac's analysis it is possible to identify the complete gauge symmetry of the theory. On the other hand,  we constructed the fundamental Dirac's brackets and we showed that the exotic action is non-commutative  and presents  problems when the cosmological constant takes the value $\Lambda=0$,  because there is  a singularity at the level of Dirac's brackets. In this respect,  we observed  an analogy with the case of Landau's problem identifying the cosmological constant  with the magnetic field and the field $e$ with the  non-commutative  coordinates. Additionally we have showed  that the exotic action is different from Palatini's theory even at the classical level; in Palatini's theory by performing a complete analysis, their Dirac's brackets  among  the dynamical variables are  commutative  and the cosmological constant can take the zero value, and there are no singularities \cite{14}. On the other hand,  we developed the canonical covariant formalism, we constructed a gauge invariant symplectic form and  we confirmed  the results obtained by means of Dirac's framework. In this manner, we have developed all tools  to  analyze the quantization aspects of the exotic action by using Dirac's canonical method or canonical covariant formalism.  \\
Finally, our analysis can be extended to others actions sharing the same equations of motion with three dimensional  gravity \cite{15}, namely
\begin{equation}
S[A, e] = S'[A,e] + \frac{1}{\gamma} \tilde{S}[A,e],
\label{eq56a}
\end{equation}
where $S'[A,e]$ is the Palatini  action, and
\begin{equation}
 \tilde{S}[A,e]= \frac{1}{\sqrt{|\Lambda |}} \left[  \frac{1}{2} \int_M A^{IJ} \wedge d A_{IJ} + \frac{2}{3}  A^{IK} \wedge  A_{KL} \wedge A^{L}{_{I}} \right] + s \sqrt{|\Lambda |}  \int_M  e_I\wedge D e^I,
 \label{eq57a}
\end{equation}
where $s$ is a constant, $\Lambda$ is the cosmological constant  and $\gamma$ is an Immirzi-like  parameter \cite{12}.  In fact, the action (\ref{eq56a}) gives rise to the same equations of motion of the Palatini action, however,  from our analysis we can observe  that Dirac's brackets of the  canonical variables $A$ and $e$ will be non-commutative. In \cite{15} was performed a canonical analysis  on a smaller phase space context of the action (\ref{eq56a}), however,  we have observed that it is mandatory to perform a detailed canonical analysis in order to know the complete symmetries. In fact,  in \cite{15}, it was not discussed the fundamental gauge symmetry of   (\ref{eq56a}),  and the case of $\Lambda=0$ was studied on a smaller phase space,  obtaining that (\ref{eq56a}) is reduced to gravity without a cosmological constant;  however, already   there exists   non-commutativity among the dynamical variables, thus,  this is not a complete study because we have commented  that  Palatini's gravity is commutative among their  dynamical variables. In this manner,  it is necessary to perform a complete Hamiltonian analysis in order to obtain  a complete description of the theory  \cite{17}. Furthermore, it is important to comment that there exist formulations of  3D gravity where has been  introduced correctly the Immirzi parameter  \cite{18, 20}.  In fact, the parameter introduced in these papers,  vanishes on half-shell, this is, when the torsion-free condition holds, which is also how the four-dimensional Immirzi parameter disappears from the Holst action. Hence, it will be useful to compare the difference among the results given in \cite{15} and those reported in \cite{18, 20}.
\newline
\newline
\noindent \textbf{Acknowledgements}\\[1ex]
This work was supported by CONACyT under Grant No. CB-2010/157641. We would like to thank   R. Cartas-Fuentevilla for discussion on the subject and   reading the manuscript.

\end{document}